# Flocculation of vesicles

Roumen Tsekov
Department of Physical Chemistry, University of Sofia, 1164 Sofia, Bulgaria

The flocculation of liposomes is theoretically studied. An expression for the flocculation activation energy is derived, accounting for the electrostatic and hydrophobic interactions as well as for the correlation area of floc-spots.

In water phospholipid molecules form spontaneously spheroid structures called vesicles or liposomes [1, 2]. Their body is filled by maternal solution and surrounded by a closed bilayer surface. The liposome wall is not penetrable for large molecules and for this reason vesicles are used as original carrier of drags and other substances in medicine and cosmetics [3, 4]. Hence, the stability of liposome suspensions is important for these applications and this is the reason for intensive studies on the flocculation kinetics of vesicles in solutions.

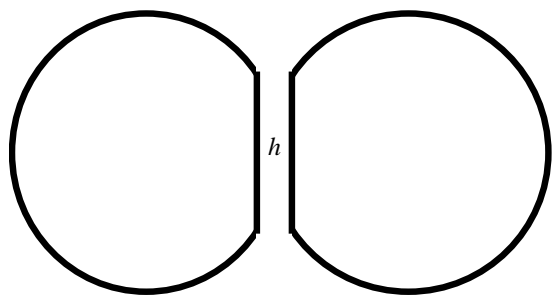

**Fig. 1** Schema of a collision of two liposomes or droplets [5]

The problem of coalescence of two liposomes is equivalent to the problem of rupture of the liquid film formed between two vesicles (see Fig. 1). Because vesicles possess almost zero surface tension, they are easily deformable and the hydrodynamic resistance force, due to the vesicle approach, deforms liposomes to form naturally a thin liquid film. Similar picture is observed in deformable droplets [5], which possess, however, large surface tension and, hence, the deformations are smaller. When the film ruptures, the liposomes attach to each other and a primary floc is formed. A very useful idea in colloid science is to consider the film stability as a result of competition of interfacial forces [6]. The classical DLVO theory considers van der Waals and electrostatic components only. It has been recognized further that there are many other macroscopic interactions in the films such as hydration, hydrophobic, etc. forces [5]. The knowledge of all these components is important to answer the questions how do the liposomes flocculate and what is the role of polymers and other substances in the floc-formation.

## Theory

Let $W$ is the free energy excess per unit area of a film as compared to the corresponding bulk liquid. The simplest expression for the van der Waals excess energy is

$$W_{VW} = -A/12\pi h^2 \qquad (1)$$

where $h$ is the film thickness and $A$ is the Hamaker constant, which is always positive for symmetric films. Hence, the van der Waals force in films dividing two liposomes or droplets is attraction. However, since the bulk phase of the vesicles possesses the same composition like the solution in the film, the Hamaker constant should be negligibly small for liposomes. Therefore, the van der Waals interaction is not important for the flocculation kinetics of vesicles, in contrast to the case of droplets.

The classical expression for electrostatic energy at constant and relatively weak surface charge $q$ per unit vesicle area is [7]

$$W_{EL} = 2(Dq^2/\varepsilon_0\varepsilon)\exp(-h/D) \qquad (2)$$

where $D$ is the Debye length and $\varepsilon_0\varepsilon$ is the dielectric permittivity of the suspension. As is seen, there are two different kinds of electrolyte effects on the electrostatic interaction. An increase of the concentration of ions not adsorbing specifically on the film interfaces decreases the Debye length and thus suppresses the electrostatic repulsion. Ions adsorbed specifically on the film interfaces change the surface charge density $q$ and depending on their sign can lead to increase or decrease of $W_{EL}$. This can be a reasonable explanation of the relative stability of liposome suspensions on ionic strength variations [8] and of the aggregation induced by calcium and beryllium cations [9].

Tsekov and Schulze have demonstrated that any adsorption of species on the film interfaces causes a hydrophobic force. The corresponding excess energy is given by [10, 11]

$$W_{HP} = 2\Delta E_G \exp(-h/a) \qquad (3)$$

where $a$ is the adsorption length and $\Delta E_G$ is the difference of the Gibbs elasticity on the surface of a vesicle and on the surface between two liposomes in a floc. The adsorption length and $\Delta E_G$ depend strongly on the concentration of species in the bulk. If $\Delta E_G$ is negative the hydrophobic force is attraction and the liposome suspensions could become unstable. Indeed, flocculation induced by addition of polymers was experimentally observed [12, 13]. Equation (3) is applicable also for description of the solvent adsorption effect. The corresponding force is known as hydration force [14] and some authors put forward an explanation of the relative stability of phospatitylcholine liposome suspensions by a strong repulsive hydration interaction [8]. The vesicle bilayers themselves are source of hydrophobic force. For instance, Alexandrova and Tsekov [15] have demonstrated that liposomes induce attractive forces in an aqueous wetting film on a quartz surface which lead to rupture of the films. The latter are stable without vesicles. Moreover, this interaction is very sensitive to small temperature changes due to many phase transitions occurring in the lipid bilyaers [16].

The total excess of free energy $W = W_{VW} + W_{EL} + W_{HP}$ in a film dividing two liposomes takes the following form

$$W = 2(Dq^2/\varepsilon_0\varepsilon)\exp(-h/D) + 2\Delta E_G \exp(-h/a) \tag{4}$$

after neglecting the contribution of the van der Waals force. Since we are interested in unstable liposome suspensions the hydrophobic force here should be attractive and $\Delta E_G$ is considered further to be negative. Usually at relatively low electrolyte concentrations the electrostatic force is longer ranged as compared to the hydrophobic one, i.e. $D \geq a$. In this case $W$ exhibits a maximum as shown in Fig. 2, if $-\varepsilon_0\varepsilon\Delta E_G \geq aq^2$. Using eq. (4) one can calculate the film thickness $h^*$ corresponding to the maximum energy and the height $W^*$ of the energy barrier

$$h^* = \frac{Da}{D-a}\ln(-\frac{\varepsilon_0\varepsilon\Delta E_G}{aq^2}) \qquad W^* = \frac{2(D-a)q^2}{\varepsilon_0\varepsilon}(-\frac{aq^2}{\varepsilon_0\varepsilon\Delta E_G})^{\frac{a}{D-a}} \tag{5}$$

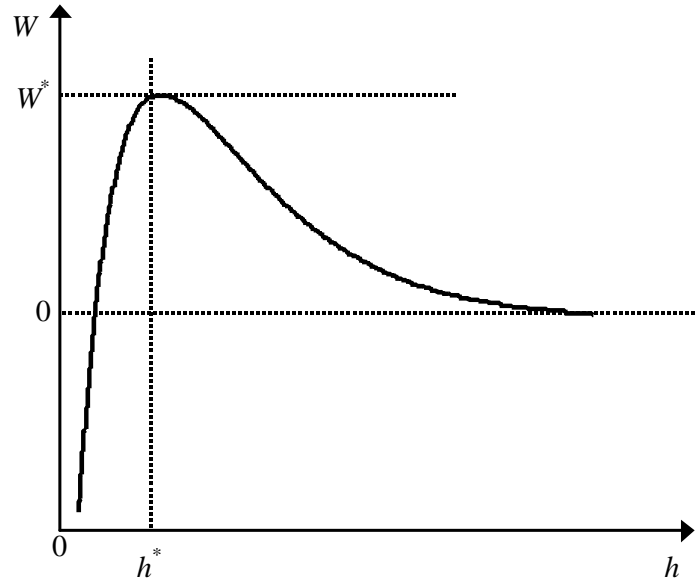

**Fig. 2** Energy profile of two liposomes separated by a film with thickness $h$

Since the pressures inside and outside a liposome are equal, the only force pushing the vesicles to each other is due to the Brownian motion and the maximum $W^*$ represents the activation energy per unit area of the liposome flocculation. As seen from eq. (5) $W^*$ is an increasing function on $D$, $q^2$ and $\Delta E_G$. The dependence $W^*$ vs. $a$ is more complicated.

The initial rate of flocculation is given by $V = K n^2$, where $n$ is the liposome concentration and $K$ is the rate constant of flocculation. Using the well-known expression

$$K = (8k_B T / 3\eta) \exp(-E^* / k_B T) \tag{6}$$

one is able to calculated the activation energy $E^*$ of the process from the temperature dependence of $K$ [17, 18]. The activation energy is related to $W^*$ and to the area $o$ of the first floc-spot formed in the film, $E^* = oW^*$. Hence, to calculate the activation energy it is necessary to calculate the correlation area $o$, as well. The shape in Fig. 1 is an average profile, while the real one fluctuates permanently. From the theory of film thickness fluctuations it is known that it is proportional to the lipid bilayer bending elasticity $\kappa$ and the second derivative of the free energy in the film in respect to its thickness. Since the fluctuations close to the barrier top are important for the flocculation kinetics the correlation area is equals to $o = \sqrt{-\kappa / W''(h^*)}$. Note that the correlation area in droplet films is equal to the square of the Scheludko wavelength, $o = -\sigma / W''(h^*)$ where $\sigma$ is the drop surface tension. Calculating the second derivative of $W$ from eq. (4) one yields $W''(h^*) = -W^* / aD$ and the correlation area amounts to

$$o = \sqrt{\kappa aD / W^*} \tag{7}$$

Finally, introducing eq. (7) in the relation $E^* = oW^*$ we acquire an expression for the activation energy

$$E^* = \sqrt{\kappa a D W^*} \tag{8}$$

It is interesting that for droplets the activation energy $E^* = \sigma aD$ is independent of $W^*$, which is due to a compensation effect via the correlation area.

The effects of other substances to the flocculation can be seen in the activation energy from eq. (8). Using eq. (5) it can be rewritten in the form

$$E^* = \sqrt{\kappa D a \frac{2(D-a)q^2}{\varepsilon_0 \varepsilon}(-\frac{aq^2}{\varepsilon_0 \varepsilon \Delta E_G})^{\frac{a}{D-a}}} \tag{9}$$

Firstly, by addition of electrolytes the Debye length $D$ decreases. This screening of the electrostatic interaction decreases the stabilising force between two vesicles. As a result the films become less stable and the activation energy $E^*$ goes down. This is the case of electrolyte induced fast coagulation in disperse systems. If some ions adsorb specifically on the liposome surface and if their charge has opposite sign than the surface one, then they diminish the surface charge density $q$ and in this way decrease the activation energy. Secondly, addition of uncharged species like polymers, aionic surfactants, etc. affects the hydrophobic force. If they decrease, the value of the resulting $\Delta E_G$ the activation energy decreases, too. A dramatic effect, however, will appear if they change also the value of the effective adsorption length. For instance, if $a$ becomes larger than $D$ there is no more maximum in the energy as a function of the film thickness. The result is barrier-less flocculation being much more rapid than the usual one. Such an increase of $a$ is expected in the case of adsorption of long polymers on the vesicle wall. And the effect is known as bridging flocculation being inter-mediated by polymer bridges. Finally, adsorption of species affects the elastic properties of the vesicle bilayers. Usually they increase the bending elastic modulus $\kappa$, which reflects in lower flocculation rate. The effect of the liposome size on flocculation can also be accounted for [19].